\documentclass{jfm}
\usepackage{amsmath,amsfonts,amssymb,bm,upmath,graphicx}
\newcommand\etal{\mbox{\textit{et al.}}}

\newcommand\eg{e.g.}

\title{Rayleigh--Taylor instability in a viscoelastic binary fluid}
\author[G. Boffetta, A. Mazzino, S. Musacchio and L. Vozella]{
 G\ls U\ls I\ls D\ls O\ns  B\ls O\ls F\ls F\ls E\ls  T\ls T\ls A$^{1,2}$\ns,
 A\ls N\ls D\ls R\ls E\ls A\ns M\ls A\ls Z\ls Z\ls I\ls N\ls O$^3$\ns, 
 S\ls T\ls E\ls F\ls A\ls N\ls O\ns M\ls U\ls S\ls A\ls C\ls C\ls H\ls I\ls 
O$^1$\ns
 and L\ls A\ls R\ls A\ns V\ls O\ls Z\ls E\ls L\ls L\ls A$^3$}
\affiliation{
 $^1$Department of General Physics, INFN and CNISM, University of Torino,\\[\affilskip]
via P. Giuria 1, 10125 Torino, Italy\\[\affilskip]
 $^2$ISAC-CNR, Sezione di Torino, 
corso Fiume 4, 10132 Torino, Italy\\[\affilskip]
 $^3$Department~of~Physics~--~University~of~Genova, and CNISM \& 
INFN~--~Unit~of~Genova, via~Dodecaneso~33, 16146~Genova, Italy}
\date{\today}

\usepackage{natbib}
\allowdisplaybreaks[1]
				   
\begin{document}

\maketitle

\begin{abstract}
The effects of polymer additives on Rayleigh--Taylor (RT) instability 
of immiscible fluids is investigated using the Oldroyd-B viscoelastic
model. 
Analytic results obtained exploiting the phase-field approach show that 
in polymer solution the growth rate of the instability speeds up with 
elasticity (but remains slower than in the pure solvent case).
Numerical simulations of the viscoelastic binary fluid model confirm this 
picture.
\end{abstract}

\section{Introduction}
Mixing of species (e.g. contaminants, tracers and particles) 
and thermodynamical quantities (e.g. temperature) are dramatically
influenced by fluid flows \cite[][]{D05}. 
Controlling the rate of mixing
in a flow is an objective of paramount importance
in many fields of science and technologies with wide-ranging consequences
in industrial applications \cite[][]{WMD01}. \\
The difficulties of the problem come from the 
intricate nature of the underlying fluid flow, 
which involves many active nonlinearly coupled 
degrees of freedom \cite[][]{F95}, and on the poor comprehension of the
way through which the fluid is coupled to the transported quantities.
The problem is even more difficult when the transported quantity reacts back
to the flow field thus affecting its dynamics. An instance is provided by the 
heat transport in convection \cite[][]{S94}.\\
Mixing emerges as a final stage of successive hydrodynamic instabilities
\cite[][]{DR81}
eventually leading to a fully developed turbulent stage. The possibility
of controlling such instability mechanisms thus allows one to have
a direct control on the mixing process. In some cases the challenge is
to enhance the mixing process by stimulating the turbulence transition,
in yet other cases the goal is to suppress deleterious instabilities 
and the ensuing turbulence.  Inertial confinement fusion \cite[][]{CZ02}
is an example whose success relies on the control of the famous 
Rayleigh--Taylor (RT) instability occurring when a heavy, denser, fluid 
is accelerated into a lighter one.
For a fluid in a gravitational field,
such instability  was first described Lord Rayleigh in the 1880s 
\cite[][]{R883}  and later generalized to all accelerated fluids by
Sir Geoffrey Taylor in 1950 \cite[][]{T50}.

Our attention here is focused on RT instability with the aim
of enhancing the perturbation growth-rate in its early stage of
evolution. The idea is to inject polymers into the fluid and to study
on both analytical and numerical 
ground how the stability of the resulting viscoelastic
fluid is modified. 
Similar problems were already investigated in more specific context,
including RT instability of viscoelastic fluids with suspended particles 
in porous medium with a magnetic field \cite[][]{SR92} and RT linear 
stability analysis of viscoelastic drops in high-speed airstream 
\cite[][]{JBF02}. We also mention that the viscoelasticity is 
known to affect also other kind of instabilities, including 
Saffman--Taylor instability \cite[][]{W90,C99},  
Faraday waves \cite[][]{WMK99,MZ99}, 
the stability of Kolmogorov flow \cite[][]{BCMPV05},
Taylor--Couette flow \cite[][]{LSM90,GS96}, 
and Rayleigh--B\'enard problem \cite[][]{VA69,ST72}.  

The paper is organized as follows. In Sec.~2 the basic equations ruling
the viscoelastic immiscible RT system are introduced 
together with the phase-field  approach. In Sec.~3 the linear analysis
is presented  and the analytical results shown and discussed 
in Sec.~4. The resulting scenario
is corroborated in Sec.~5 by means of direct numerical simulations of the 
original field equations.

\section{Governing equations}
The system  we consider is composed of two incompressible fluids 
(labeled by 1 and 2) having different densities, 
$\rho_1$ and $\rho_2 > \rho_1$, and different dynamical viscosities, 
$\mu_1$ and $\mu_2$, with the denser fluid placed above 
the less dense one. For more generality, the two fluids are supposed 
to be immiscible so that the surface tension on the interface separating 
the two fluids will be explicitly taken into account. 

The effects of polymer additives is here studied within the framework of the 
Oldroyd-B model \cite[][]{O50,H77,BHAC87}.
In this model polymers are treated as elastic
dumbbells, i.e. identical pairs of microscopic beads connected by harmonic 
springs. Their concentration is supposed to be  
low enough to neglect polymer-polymer interactions. 
The polymer solution is then regarded as a continuous medium, 
in which the reaction of polymers on the flow is 
described as an elastic contribution to the 
total stress tensor of the fluid \cite[see \eg ][]{BHAC87}.

In order to describe the mixing process of the resulting viscoelastic
immiscible fluids we follow the phase-field approach 
(for a general description of the method see, \eg ,~\cite{B02, CH58}, 
and for application 
to multiphase flows see, \eg ,~\cite{BCB03,DSS07,M07,CMMV09}). 
Here, we only recall that
the basic idea of the method is to treat
the interface between two immiscible fluids as a thin mixing layer 
across which physical properties vary steeply but continuously. 
The evolution of the mixing layer is ruled by an order parameter 
(the phase field)  that obeys a Cahn--Hilliard equation \cite[][]{CH58}. 
One of the advantage of the method is that the boundary conditions at
the fluids interface need not to be specified being encoded in the
governing equations.
From a numerical point of view, the method permits to 
avoid a direct tracking of the interface and easily 
produces the correct interfacial tension from the mixing-layer free energy.

To be more specific, the evolution of the viscoelastic binary fluid 
is described by the system of differential equations
\begin{equation}
\rho_0 
\left(\partial_t {\bm v} + {\bm v} \cdot{\bm \partial} {\bm v}\right) = 
-{\bm \partial} p + {\bm \partial} \cdot (2 \mu {\bm e} )
+ A \rho_0 {\bm g} \phi  
-\phi {\bm \partial} {\cal M}
+{2 \mu \eta \over \tau}
{\bm\partial}\cdot ({\bm \sigma}-\mathbb{I})
\label{eq1} 
\end{equation}
\begin{equation}
\partial_t \phi + {\bm v} \cdot {\bm \partial} \phi = \gamma
\partial^2 {\cal M}
\label{eq2} 
\end{equation}
\begin{equation}
\partial_t {\bm \sigma}+{\bm v} \cdot{\bm \partial} {\bm \sigma} = 
({\bm \partial}{\bm v})^T \cdot{\bm \sigma}+ 
{\bm \sigma}\cdot{\bm \partial}{\bm v}-{2 \over \tau}({\bm \sigma}- 
\mathbb{I})
\,\,\, .
\label{eq3} 
\end{equation}
Eq.~(\ref{eq1}) is the usual Boussinesq Navier--Stokes equation 
\cite[][]{KC01} with two additional stress contributions. 
The first one arises 
at the interface where 
the effect of surface tension enters into play \cite[][]{B02,YFLS04,BBCV05},
the last term represents the polymer back-reaction to the flow field
\cite[][]{BHAC87}.\\
In (\ref{eq1}),
we have defined $\rho_0=(\rho_1 + \rho_2)/2$, $\bf{g}$ is the gravitational 
acceleration pointing along the $y$-axis, 
$\mathcal{A}\equiv(\rho_2-\rho_1)/(\rho_2+\rho_1)$ 
is the Atwood number, $e_{ij}\equiv\left(
\partial_i v_j + \partial_j v_i \right) / 2$ is the rate of strain tensor and
$\mu=\mu(\phi)$ is the dynamical viscosity field parametrically defined as 
\cite[][]{LS03}
\begin{equation}
\frac{1}{\mu} = \frac{1+\phi}{2 \mu_1}+\frac{1-\phi}{2 \mu_2} 
\label{eq4}
\end{equation}
$\phi$ being the phase field governed by (\ref{eq2}).  
The phase field $\phi$ is representative of density fluctuations and 
we take $\phi=1$ in the regions of density $\rho_1$ 
and $\phi=-1$ in those of density $\rho_2 \ge \rho_1$. 
${\bm \sigma}\equiv \frac{\langle{\bm R}{\bm R} \rangle}{R_0^2}$ 
is the polymer conformation tensor, ${\bm R}$ being the end-to-end  
polymer vector ($R_0$ is the polymer length at equilibrium), 
the parameter $\eta$ is proportional to polymer concentration and
$\tau=\tau(\phi)$ is the (slowest) polymer relaxation time which, according
to the Zimm model \cite[][]{DE86}, is assumed to be proportional to 
the viscosity $\mu$ (therefore we have $\tau=\tau_1$ for $\phi=1$ and
$\tau=\tau_2$ for $\phi=-1$ with $\mu(\phi)/\tau(\phi)$ constant).
Finally, $\gamma$ is the mobility and ${\cal M}$ is the chemical 
potential defined in terms of 
the Ginzburg--Landau free energy ${\cal F}$ as \cite[][]{CH58,B02,YFLS04} 
\begin{equation}
{\cal M} \equiv \frac{\delta {\cal F} }{\delta \phi}\qquad\mbox{and}\qquad
{\cal F}[\phi] \equiv \lambda \int_{\Omega} \mathrm{d}\bm{x} \;
\left( \frac{1}{2} |{\bm \partial} \phi|^2+ V(\phi) \right) \, . 
\label{eq5}
\end{equation}
where $\Omega$ is the region of space occupied by the system, $\lambda$ 
is the magnitude of the free-energy and the potential $V(\phi)$ is 
\begin{equation}
V(\phi)\equiv \frac{1}{4 \epsilon^2} (\phi^2 -1 )^2
\label{eq6}
\end{equation}
where $\epsilon$ is the capillary 
width, representative of the interface thickness.\\

The unstable equilibrium state with heavy fluid placed on the top of
light fluid is given by
\begin{equation}
\bm{v}=\bm{0}\, , \quad \phi(y)=-\tanh\left (\frac{y}
{\epsilon\sqrt{2}}\right )\qquad \mbox{and}\qquad {\bm \sigma}=\mathbb{I}
\label{eq7}
\end{equation}
corresponding to a planar interface of  width 
$\epsilon$ with polymers having their equilibrium length $R_0$.  
In this case,  the surface tension,  ${\cal S}$, is given by
\cite[see, for example,][]{LL00}:
\begin{equation}
{\cal S} \equiv \lambda \int_{-\infty}^{+\infty} dy \;\left( \frac{1}{2} 
|{\bm \partial} \phi|^2+ V(\phi)
\right) = \frac{2\lambda \sqrt{2}}{3\epsilon} \, .
\label{eq8}
\end{equation}
The sharp-interface limit is obtained by taking
the $\lambda$ and $\epsilon$ to zero,  
keeping $\cal{S}$ fixed to the value prescribed by surface tension  
 \cite[][]{LS03}. 

\section{Linear stability analysis}
Let us now suppose to impose a small perturbation on the 
interface separating the two  fluids. 
Such perturbation will displace the phase field from 
the previous equilibrium configuration, which minimizes the free energy 
(\ref{eq5}) to a new configuration for which, in general,  
$\mathcal{M} \neq 0$. We want to determine how the perturbation evolves 
in time.

Focusing on the two-dimensional case (corresponding to translational invariant 
perturbations along the $z$ direction), let us
denote by $h(x,t)$ the perturbation imposed to the planar interface
$y=0$ in a way that we can rewrite the phase-field $\phi$ as:
\begin{equation}
\phi = f\left(\frac{y-h(x,t)}{\epsilon \sqrt{2}}\right)\, ,
\label{eq9}
\end{equation}
where $h$ can be larger than $\epsilon$, yet it has to
be smaller than the scale of variation of $h$ (small amplitudes).
In this limit we assume the interface to be locally in equilibrium,
i.e. $\partial^2 f/\partial y^2 = V'(f)$, and thus $f(y)=-\tanh(y)$
and therefore ${\cal M} = - \lambda \frac{\partial^2 f}{\partial x^2}$
($'$ denotes derivative with respect to the argument).

Linearizing the momentum equation for small interface
velocity we have
\begin{equation}
\rho_0 \partial_t v_y = - \partial_y p - \phi \partial_y {\cal M} - 
A g \rho_0 \phi + {2 \mu \eta \over \tau}\partial_i\sigma_{i2}
+\mu \left(\partial_x^2 + \partial_y^2 \right)v_y
+2(\partial_y v_y)\partial _y\mu \,. 
\label{eq10}
\end{equation}

Integrating on the vertical direction and using derivations by parts 
one gets
\begin{equation}
\rho_0 \partial_t q = {\cal S} \frac{\partial^2 h}{\partial x^2}+ 
2 A g \rho_0 h + {2 \mu \eta \over \tau} \Sigma +Q
\label{eq11}
\end{equation}
where we have defined 
\begin{equation}
Q\equiv \int_{-\infty}^{+\infty}\mu 
\left (\frac{\partial^2 }{\partial x^2} -
\frac{\partial^2 }{\partial y^2}\right )v_y\,dy\qquad
q \equiv  \int_{-\infty}^{\infty} v_y\,dy  \qquad
\Sigma\equiv \int_{-\infty}^{\infty}\partial_x \sigma_{12}\,dy \, ,
\label{eq12}
\end{equation}
and we have used the relations
$\int (f')^2 dy = 2 \sqrt{2}/(3 \epsilon)$, 
$\int f f''' dy = 0$, $\int f dy = 2 h$.

Note that, unlike what happens in the inviscid case,  
Eq.~(\ref{eq11}) does not involve solely
the field $q_y$ but also second-order derivatives of $v_y$.
In order to close the equation,  
let us resort to a potential-flow description.
The idea is to evaluate $Q$ for a potential flow $v_y$ and then to 
plug $Q=Q^{pot}$  into (\ref{eq11}) \cite[][]{M93}.
The approximation is justified when viscosity is sufficiently small 
and its effects are confined in a narrow region around the interface. 
Because for a potential flow $\partial^2 {\bm v}=0$ we have
\begin{equation} 
Q^{pot}=
2 \int_{-\infty}^{+\infty}\mu {\partial^2 u_y \over \partial x^2}\, dy=
2 \int_{-\infty}^{0}\mu {\partial^2 u_y \over \partial x^2}\, dy +
2 \int_{0}^{\infty}\mu {\partial^2 u_y \over \partial x^2}\, dy =
(\mu_1+\mu_2){\partial^2 q \over \partial x^2} \, .
\label{eq14}
\end{equation}
Substituting in (\ref{eq11}) and defining $\nu=(\mu_1+\mu_2)/(2 \rho_0)$
one finally obtains
\begin{equation}
\partial_t q =
{{\cal S} \over \rho_0} {\partial^2 h \over \partial x^2}+ 2 A g h +  
{2 \mu \eta \over \tau \rho_0} \Sigma + 
2 \nu {\partial^2 q \over \partial x^2} \, .
\label{eq15}
\end{equation}
Let us now exploit the equation (\ref{eq2}) for the phase field to relate
$q_y$ to $h$. For small amplitudes, we have:
\begin{equation}
\partial^2{\cal M} = {\lambda \over \epsilon \sqrt{2}}
\left [f'\frac{\partial^4h}{\partial x^4} + {1 \over 2\epsilon^2} 
f'''\frac{\partial^2 h}{\partial x^2} \right ]
\label{eq16}
\end{equation}
and therefore, from (\ref{eq2})
\begin{equation}
- {1 \over \epsilon} f' \partial_t h + v_y {1 \over \epsilon} f' = 
{\gamma \lambda \over \epsilon} \left [f' \partial_x ^4 h + 
{1 \over 2\epsilon^2} f''' \partial_x^2 h \right ] \, .
\label{eq17}
\end{equation}

Integrating over $y$, observing that $1/(2\sqrt{2}\epsilon) f'$
approaches $\delta(y-h)$ as $\epsilon \to 0$ and using the limit
of sharp interface ($\gamma \lambda \to 0$) one obtains
\begin{equation}
\partial_t h = v_y(x,h(t,x),t) \equiv v_y^{(int)}(x,t) \, .
\label{eq18}
\end{equation}

The equation for the perturbation $\sigma_{12}$ of the conformation 
tensor is obtained by linearizing (\ref{eq3}) around 
$\sigma_{\alpha \beta}=\delta_{\alpha \beta}$
\begin{equation}
\partial_t \sigma_{12}= \partial_{x} v_{y}+ \partial_{y} v_{x} -
\frac{2}{\tau}\sigma_{12}
\label{eq19}
\end{equation}
from which, exploiting incompressibility, we obtain
\begin{equation}
\partial_t \partial_x \sigma_{12}= 
(\partial_{x}^2 - \partial_{y}^2) v_{y} -
{2  \over \tau} \partial_x \sigma_{12} - 
2 \sigma_{12} \partial_x {1 \over \tau}\, .
\label{eq20}
\end{equation}
For small amplitude perturbations the last term, which is proportional to 
$\sigma_{12} \partial_x \phi$, can be neglected at the leading order.
Integrating over $y$ and using again the potential flow approximation
one ends up with 
\begin{equation}
\partial_t \Sigma=2 \partial_x^2 q - {2 \over \bar{\tau}} \Sigma 
- ({1 \over \tau_1}-{1 \over \tau_2}) \int dy \phi \partial_x \sigma_{12}
\, .
\label{eq21}
\end{equation}
where we have introduced $\bar{\tau}=2 \tau_1 \tau_2/(\tau_1 + \tau_2)$.

In conclusion, we have the following set of equations (in the $(x,t)$ 
variables) for the
linear evolution of the Rayleigh--Taylor instability in a 
viscoelastic flow
\begin{equation}
\left \{
\begin{array}{lll}
\partial_t h & = & v_y^{(int)} \\
\partial_t q & = & 
{{\cal S} \over \rho_0} \partial_x^2 h + 2 A g h +  
{2 \nu \eta c \over \bar{\tau}} \Sigma + 2 \nu \partial_x^2 q  \\
\partial_t \Sigma & = & 2 \partial_x^2 q - {2 \over \bar{\tau}} \Sigma 
- ({1 \over \tau_1}-{1 \over \tau_2}) \int dy \phi \partial_x \sigma_{12}
\, .
\end{array}
\right .
\label{eq22}
\end{equation}
where $c=4 \mu_1 \mu_2/(\mu_1+\mu_2)^2 \le 1$.

\section{Potential flow closure for the interface velocity}
The set of equations (\ref{eq22}) is not closed because
of the presence of the interface velocity $v_y^{(int)}$
and of the integral term in the equation for $\Sigma$.
In order to close the system we exploit again the potential
flow approximation for which $v_y=\partial_y \psi$. 

Taking into account the boundary condition for $y\to \infty$, the potential 
can be written (e.g.~for $y\ge 0$) as
\begin{equation}
\psi(x,y,t)=\int_0^{\infty} e^{-k y+i k x} \hat{\psi}(k,t) dk + c.c.
\label{eq23}
\end{equation}
where ``\^{ }'' denotes the Fourier transform, and therefore
\begin{equation}
v_y(x,y,t)= - \int_0^{\infty} k e^{-k y+i k x} \hat{\psi}(k,t) dk + c.c.
\label{eq24}
\end{equation}
\begin{equation}
q(x,t)= - 2 \int_0^{\infty} e^{i k x} \hat{\psi}(k,t) dk + c.c.
\label{eq25}
\end{equation}
and taking a flat interface, $y=0$, at the leading order
\begin{equation}
v^{(int)}(x,t)= - \int_0^{\infty} k e^{i k x} \hat{\psi}(k,t) dk + c.c.
\label{eq26}
\end{equation}
Assuming consistently that also 
\begin{equation}
\sigma_{12}(x,y,t)= \int_0^{\infty} e^{-k y+i k x} \hat{\sigma}_{12}(k,t) dk + c.c. \, ,
\label{eq26b}
\end{equation}
in the limit of small amplitudes one has 
$\int dy \phi \partial_x \sigma_{12}=0$ and
the set of equation (\ref{eq22}) for the Fourier coefficients becomes
\begin{equation}
\left \{
\begin{array}{lll}
\partial_t \hat{h} & = & {k \over 2} \hat{q} \\
\partial_t \hat{q} & = & 
- {{\cal S} \over \rho_0} k^2 \hat{h} + 2 A g \hat{h} +  
{2 \nu c \eta \over \bar{\tau}} \hat{\Sigma} - 2 \nu k^2 \hat{q}  \\
\partial_t \hat{\Sigma} & = & - 2 k^2 q - {2 \over \bar{\tau}} \hat{\Sigma} \, .
\end{array}
\right .
\label{eq27}
\end{equation}
Restricting first to the case without polymers ($\eta=0$), 
the growth rate $\alpha_N$ 
of the perturbation is obtained by looking for a solution
of the form $\hat{h} \sim e^{\alpha_N t}$ which gives
\begin{equation}
\alpha_N = - \nu k^2 + \sqrt{\omega^2 + (\nu k^2)^2}
\label{eq28}
\end{equation}
where it has been defined
\begin{equation}
\omega=\sqrt{A g k - {{\cal S} \over 2 \rho_0} k^3} \, .
\label{eq29}
\end{equation}
The expression (\ref{eq29}) is the well-known growth rate for a 
Newtonian fluid in the limit of zero viscosity \cite[][]{C61},
while (\ref{eq28}) is a known upper bound to the growth rate for 
the case with finite viscosity \cite[][]{MMSZ77}. 

Let us now consider the case with polymers, i.e. $\eta>0$.
The growth rate $\alpha$ is given by the solution of
\begin{equation}
(\alpha \bar{\tau})^3 + 2 (\alpha \bar{\tau})^2 (1+\nu k^2 \bar{\tau})+\alpha 
\bar{\tau}
\left[4 \nu (1+c \eta) k^2 \bar{\tau} -\omega^2 \bar{\tau}^2 \right]-2 \omega^2 \bar{\tau}^2=0 \, .
\label{eq30}
\end{equation}
The general solution is rather complicated and not very enlightening.
In the limit of stiff polymers, $\bar{\tau} \to 0$, one gets
\begin{equation}
\alpha_0 \equiv \lim_{\bar{\tau} \to 0} \alpha = - \nu(1+c \eta) k^2 + \sqrt{\omega^2 + [\nu(1+c \eta) k^2]^2} \, .
\label{eq31}
\end{equation}
Comparing with (\ref{eq28}) one sees that in this limit polymers 
simply renormalize solvent viscosity. This result is in agreement
with the phenomenological definition of $c \eta$ as the zero-shear
polymer contribution to the total viscosity of the mixture \cite[][]{V75}.
Therefore, in order to quantify the effects of elasticity on RT instability,
the growth rate for viscoelastic cases at finite $\bar{\tau}$ 
has to be compared with the Newtonian case with renormalized
viscosity $\nu(1+c \eta)$. 

Another interesting limit is $\bar{\tau} \to \infty$. In this case from 
(\ref{eq30}) one easily obtains that the growth rate 
coincides with that of the 
pure solvent (\ref{eq28}), i.e. $\alpha_{\infty} = \alpha_N$.
The physical interpretation is that in the limit $\bar{\tau} \to \infty$
and at finite time for which polymer elongation is finite, the last
term in (\ref{eq1}) vanishes and one recovers the Newtonian 
case without polymers (i.e. $\eta=0$).
Of course, this does not mean that in general polymer effects 
for high elasticity disappear. 
Indeed in the long-time limit polymer elongation is able to
compensate the $1/\tau$ coefficient and in the late, non-linear
stages, one expects to observe strong polymer effects at high elasticity.

From equation (\ref{eq30}) one can easily show (using implicit differentiation)
that $\alpha(\bar{\tau})$ is a monotonic  function and, because
$\alpha_\infty \ge \alpha_0$, we have that instability rate
grows with the elasticity, or the Deborah number, here defined
as $De \equiv \omega \bar{\tau}$.

The case of stable stratification, $g \to -g$, is obtained 
by $\omega^2 \to -\omega^2$ neglecting surface tension.
In this case (\ref{eq30}) has no solution for positive $\alpha$,
therefore polymers alone cannot induce instabilities in a stably 
stratified fluid.

\section{Numerical results}
The analytical results obtained in the previous Sections
are not exact as they are based on a closure obtained
from the potential flow approximation. While this approximation is consistent 
for the inviscid limit $\nu=0$ (where it gives the 
correct result (\ref{eq29}) for a Newtonian fluid) for finite 
viscosity we have shown
that it gives a known upper bound to the actual growth rate of
the perturbation \cite[][]{MMSZ77} (this is because the potential
flow approximation underestimates the role of viscosity which
reduces the instability). Nonetheless, in the case of 
Newtonian fluid this upper bound is known to be a good approximation 
of the actual  value of the growth rate measured in
numerical simulations \cite[][]{MMSZ77}.
Because both $\bar{\tau} \to 0$ and $\bar{\tau} \to \infty$ limits 
correspond to Newtonian fluids, we expect that also in 
the viscoelastic case the potential flow description is a good approximation.

To investigate this important point, we have performed a
set of numerical simulations of the full model (\ref{eq1}-\ref{eq3})
in the limit of constant viscosity and relaxation time 
(i.e. $\mu_1=\mu_2$, $c=1$ and $\tau_1=\tau_2=\bar{\tau}$)
in two dimensions by means of a standard, fully dealiased,
pseudospectral method on a square doubly periodic domain.
The resolution of the simulations is $1024\times 1024$ collocation points
(a comparative run at double resolution 
did not show substantial modifications on the results).
More details on the numerical simulation method can be found
in \cite{CMV06} and \cite{CMMV09}.

The basic state corresponds to a zero velocity field, 
a hyperbolic-tangent profile for the phase field and an uniform 
distribution of polymers in equilibrium,
according to  (\ref{eq7}).
The interface of the basic state is perturbed
with a sinusoidal wave at wavenumber $k$ (corresponding to 
maximal instability for the linear analysis)
of amplitude $h_0$ much smaller than the wavelength 
($k h_0 =0.05$).

\begin{figure}
\centering
\includegraphics[scale=0.8]{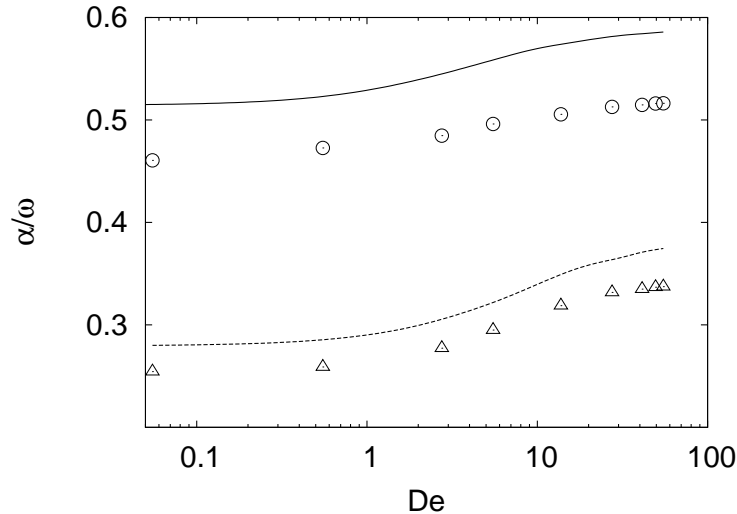} 
\caption{The perturbation growth-rate $\alpha$ normalized
with the inviscid growth rate $\omega$ (\ref{eq29}) as a 
function of the Deborah number $De=\omega \bar{\tau}$.
Points are the results of numerical simulations of the
full set of equations (\ref{eq1}-\ref{eq3}), lines represent
the theoretical predictions obtained from (\ref{eq30}).
The values of parameters are: $c=1$, $k=1$, $A g=0.31$, 
$\mathcal{S} / \rho_0=0.019$ and $\eta=0.3$, $\nu=0.3$
(upper points and line) and $\eta=0.5$, $\nu=0.6$ (lower
points and line).}
\label{fig1}
\end{figure}

The growth rate $\alpha$ of the perturbation is measured
directly by fitting the height of the perturbed interface at
different times with an exponential law. 
For given values of $A\,g$, $\mathcal{S}/ \rho_0$, $\nu$ and $\eta$, 
this procedure is repeated for different values of $\bar{\tau}$ at the 
maximal instability wavenumber $k$ (which, for
the range of parameters considered here, is always $k=1$, i.e. 
it is not affected by elasticity).
Figure~\ref{fig1} shows the results for two sets of runs at
different values of $\eta$ and $\nu$.
As discussed above, we find that
the theoretical prediction given by (\ref{eq30}) is indeed an upper 
bound for the actual growth rate of the perturbation. 
Nevertheless, the bound gives grow rates which are quite close
to the numerical estimated values (the error is of the order of 
$10\%$).
The error is smaller for the runs having a larger value of $\eta$ and
$\nu$, as was already discussed by \cite{CMMV09}.

Both theoretical and numerical results show that the effect of 
polymers is to increase the perturbation growth-rate. 
$\alpha$ grows with the elasticity and saturates for sufficiently large value 
of $De$.  

\section{Conclusions and perspectives}
We investigated the role of polymers on the linear phase of the 
Rayleigh--Taylor instability in an Oldroyd-B viscoelastic model.
In the limit of vanishing Deborah number (i.e.
vanishing polymer relaxation time) we recover a known
upper bound for the growth rate of the perturbation in a
viscous Newtonian fluid with modified viscosity. 
For finite elasticity,
the growth rate is found to increase monotonically with 
the Deborah number reaching the solvent limit for high Deborah numbers.
Our findings are corroborated by a set of direct numerical simulations on the
viscoelastic Boussinesq Oldroyd-B model.

Our analysis has been confined to the linear phase of the perturbation 
evolution. When the perturbation amplitude becomes sufficiently large, 
nonlinear effects enter into play and a fully developed turbulent regime
rapidly sets in \cite[][]{CC06,VC09,BMMV09}.
In the turbulent stage we expect more dramatic effects of polymers. 
In turbulent flows, 
a spectacular consequence of viscoelasticity induced by polymers 
is the drag reduction effect: addition 
of minute amounts (a few tenths of p.p.m. in weight) 
of long-chain soluble polymers 
to water leads to a strong reduction (up to $80\%$) 
of the power necessary to maintain a 
given throughput in a channel \cite[see \eg ][]{T49,V75}.
We conjecture that a similar phenomenon
might arise also in the present context. 
Heuristically, the RT system can indeed be 
assimilated to a channel inside
which vertical motion of thermal plumes 
is maintained by the available potential energy.
This analogy suggests the possibility 
to observe in the viscoelastic RT system a ``drag'' reduction
(or mixing enhancement) phenomenon, 
i.e. an increase of the velocity of thermal plumes with respect to the
Newtonian case. 
Whether or not this picture does apply 
to the fully developed turbulence regime
is left for future research. 

We thank anonymous Referees for useful remarks.

\end{document}